\documentclass[PRC,aps,noshowpacs,twocolumn,superscriptaddress,floatfix]{revtex4-1}

\usepackage{ulem}

\usepackage[dvipdfmx]{graphicx}
\usepackage{here}
\usepackage{amsmath,amssymb,times}
\usepackage{color}

\newcommand{\beq}{\begin{equation}}
\newcommand{\eeq}{\end{equation}}
\newcommand{\bea}{\begin{eqnarray}}
\newcommand{\eea}{\end{eqnarray}}

\newcommand{\bfi}[1]{\mbox{\boldmath $#1$}}

\newcommand{\vK}{{\bfi K}}

\newcommand{\vs}{{\bfi s}}

\newcommand{\vrr}{{\bfi r}}
\newcommand{\vR}{{\bfi R}}

\def\a{\alpha}

\begin{document}

\title{Reanalyses for  $^{42-51}$Ca scattering on a $^{12}$C target at $280$~MeV/nucleon \\ 
based on chiral $g$ folding mode with Gogny-D1S Hartree-Fock-Bogoliubov densities 
}

\author{Maya~Takechi}
\affiliation{Niigata University, Niigata 950-2181, Japan}

\author{Tomotsugu~Wakasa}
\affiliation{Department of Physics, Kyushu University, Fukuoka 819-0395, Japan}

\author{Shingo~Tagami}
\affiliation{Department of Physics, Kyushu University, Fukuoka 819-0395, Japan}

\author{Jun~Matsui}
\affiliation{Department of Physics, Kyushu University, Fukuoka 819-0395, Japan}

\author{Masanobu~Yahiro}
\email[]{orion093g@gmail.com}
\affiliation{Department of Physics, Kyushu University, Fukuoka 819-0395, Japan}

\begin{abstract}
\begin{description}
\item[Background]  In the previous paper, we predicted reaction cross sections $\sigma_{\rm R}$ for $^{40-60,62,64}$Ca+$^{12}$C scattering at $280$~MeV/nucleon, using the chiral $g$-matrix folding model with the densities 
calculated with the Gogny-D1S Hartree-Fock-Bogoliubov (GHFB) with and without the angular momentum projection 
(AMP), since Tanaka {\it el al.} measured interaction cross sections $\sigma_{\rm I} (\approx \sigma_{\rm R})$ 
for $^{42-51}$Ca in RIKEN and determined neutron skin $r_{\rm skin}({\rm RIKEN})$ 
using the optical limit of the Glauber model with the Woos-Saxon densities. 
\item[Purpose]     Our purpose is to reanalyze the $r_{\rm skin}$ from the $\sigma_{\rm I}$ 
using the chiral $g$-matrix folding model. 
Our analysis is superior to theirs, since the chiral $g$-matrix folding model (the  GHFB and GHFB+AMP densities) is much better than the optical limit of the Glauber model (the Woos-Saxon densities). 
\item[Methods]     Our model is the chiral $g$-matrix folding model with the densities scaled from the GHFB and GHFB+AMP densities. 
\item[Results]     We scale the  GHFB and GHFB+AMP densities
so that the $\sigma_{\rm R}$ of the scaled densities can agree with the central values of $\sigma_{\rm I}$ 
under the condition that the proton radius of the scaled proton density 
equals the data determined from the isotope shift  based on the electron scattering. 
The $r_{\rm skin}$ thus determined  are close to their results $r_{\rm skin}^{42-51}({\rm RIKEN})$, 
except for $^{48}$Ca.  
For $^{48}$Ca, our value $r_{\rm skin}^{48}$ is 0.105 $\pm$ 0.06~fm, while their value is 
$r_{\rm m}^{48}({\rm RIKEN})=0.146 \pm 0.06$~fm. 
We then take the weighted mean and its error of 
our result   $r_{\rm skin}^{48}(\sigma_{\rm I})= 0.105 \pm 0.06$~fm and 
the result $r_{\rm skin}^{48}(E1{\rm pE}) =0.17 \pm 0.03$~fm of the high-resolution $E1$ polarizability experiment (E1{\rm pE}). 
Our final result is $r_{\rm skin}^{48}=0.157 \pm 0.027$~fm. 
\item[Conclusion] 
Our conclusion is $r_{\rm skin}^{48}=0.157 \pm 0.027$~fm for $^{48}$Ca.  
For $^{42-47,49-51}$Ca, our results on $r_{\rm skin}$ are similar to theirs. Our result for  $^{48}$Ca is related to CREX.
\end{description}
\end{abstract}

\maketitle

\section{Introduction}

Very lately, Tanaka {\it el al.} measured interaction cross sections 
$\sigma_{\rm I}$ in RIKEN for $^{42-51}$Ca+ $^{12}$C scattering 
at 280~MeV per nucleon, and 
determined neutron skins $r_{\rm skin}$ for $^{42-51}$Ca from the $\sigma_{\rm I}$, 
using the optical limit of the Glauber model with the Woos-Saxon densities~\cite{Tanaka:2019pdo}. 
The data have high accuracy, since the average error is 1.1\%. 
Their numerical values on matter radii $r_{\rm m}({\rm RIKEN}) $, skin values $r_{\rm skin}({\rm RIKEN})$, 
neutron radii $r_{\rm n}({\rm RIKEN})$,  determined from  $\sigma_{\rm I}$ are not presented
 in Ref.~\cite{Tanaka:2019pdo}; see Table  \ref{Radii-exp} for their numerical values.

\begin{table}[H]
\begin{center}
\caption
{Numerical values~\cite{Tanaka:2019pdo} 
of $r_{\rm m}({\rm RIKEN})$, $r_{\rm n}({\rm RIKEN})$, $r_{\rm skin}({\rm RIKEN})$ for $^{42-51}$Ca. 
The $r_{\rm p}({\rm exp})$ are deduced from the electron scattering~\cite{Angeli:2013epw}.
The radii are shown in units of fm. The errors include systematic errors. 
 }
\begin{tabular}{ccccc}
\hline\hline
A & $r_{\rm p}({\rm exp})$ & $r_{\rm m}({\rm RIKEN})$ &  $r_{\rm n}({\rm RIKEN})$ & $r_{\rm skin}({\rm RIKEN})$ \\
\hline
42 & $3.411 \pm 0.003$ & $3.437 \pm 0.030$ & $3.46\pm 0.06$ & $0.049\pm 0.06$ \\
43 & 3.397 $\pm$ 0.003& 3.453 $\pm$ 0.029 & 3.50 $\pm$ 0.05 & 0.103 $\pm$ 0.05\\
44 & 3.424 $\pm$ 0.003 & 3.492 $\pm$ 0.030 & 3.55 $\pm$ 0.05 & 0.125 $\pm$ 0.05\\
45 & 3.401 $\pm$ 0.003 & 3.452 $\pm$ 0.026 & 3.49 $\pm$ 0.05 & 0.092 $\pm$ 0.05\\
46 & 3.401 $\pm$ 0.003 & 3.487 $\pm$ 0.026 & 3.55 $\pm$ 0.05 & 0.151 $\pm$ 0.05\\
47 & 3.384 $\pm$ 0.003 & 3.491 $\pm$ 0.034 & 3.57 $\pm$ 0.06  & 0.184 $\pm$ 0.06\\
48 & 3.385 $\pm$ 0.003 & 3.471 $\pm$ 0.035 & 3.53 $\pm$ 0.06 & 0.146 $\pm$ 0.06\\
49 & 3.400 $\pm$ 0.003 & 3.565 $\pm$ 0.028 & 3.68 $\pm$ 0.05 & 0.275 $\pm$ 0.05\\
50 & 3.429 $\pm$ 0.003 & 3.645 $\pm$ 0.031 & 3.78 $\pm$ 0.05  & 0.353 $\pm$ 0.05\\
51 & 3.445 $\pm$ 0.003 & 3.692 $\pm$ 0.066 & 3.84 $\pm$ 0.10 & 0.399 $\pm$ 0.10\\
\hline
\end{tabular}
 \label{Radii-exp}
 \end{center} \end{table}

The $g$-matrix folding model~\cite{Brieva-Rook,Amos,CEG07,Minomo:2011bb, Sumi:2012fr, Egashira:2014zda,Watanabe:2014zea,Toyokawa:2013uua,Toyokawa:2014yma,Toyokawa:2015zxa,Toyokawa:2017pdd} is a standard way of 
determining matter radii  $r_{\rm m}$ from measured reaction cross sections $\sigma_{\rm R}$. 
In the model, the  potential is obtained by folding  the $g$-matrix with projectile and target densities.

Applying the Melbourne $g$-matrix folding model~\cite{Amos} for  interaction cross sections $\sigma_{\rm I}$ 
of Ne isotopes and 
reaction cross sections $\sigma_{\rm R}$ of Mg isotopes, we 
deduced the $r_{\rm m}$ for Ne isotopes~\cite{Sumi:2012fr} and  
Mg isotopes~\cite{Watanabe:2014zea}, and discovered 
that $^{31}$Ne is a halo nucleus with large deformation~\cite{Minomo:2011bb}. 

Kohno calculated the $g$ matrix  for the symmetric nuclear matter, 
using the Brueckner-Hartree-Fock method with chiral N$^{3}$LO 2NFs and NNLO 3NFs~\cite{Koh13}. 
He set $c_D=-2.5$ and $c_E=0.25$ so that  the energy per nucleon can  become minimum 
at $\rho = \rho_{0}$~\cite{Toyokawa:2017pdd}.

Toyokawa {\it et al.} localized the non-local chiral  $g$ matrix 
into three-range Gaussian forms by using the localization method proposed 
by the Melbourne group~\cite{von-Geramb-1991,Amos-1994,Amos}. 
The resulting local  $g$ matrix is called  ``Kyushu  $g$-matrix''; 
see the hompage  http://www.nt.phys.kyushu-u.ac.jp/english/gmatrix.html for Kyushu  $g$-matrix.

The  Kyushu $g$-matrix folding model is successful in reproducing $d\sigma/d\Omega$ and $A_y$
for polarized  proton scattering  on various targets at $E_{\rm lab}=65$~MeV~\cite{Toyokawa:2014yma}
and $d\sigma/d\Omega$ for $^4$He scattering at $E_{\rm lab}=72$~MeV per nucleon~\cite{Toyokawa:2015zxa}. 
This is true for $\sigma_{\rm R}$ of $^4$He scattering 
in $E_{\rm lab}=30 \sim 200$~MeV per nucleon~\cite{Toyokawa:2017pdd}. 

In the previous paper of Ref.~\cite{Tagami:2019svt},  
we predicted reaction cross section $\sigma_{\rm R}$ for $^{40-60,62,64}$Ca  scattering on a $^{12}$C target at $280$~MeV/nucleon, using the Kyushu $g$-matrix folding model with the reliable densities 
calculated with the Gogny-D1S Hartree-Fock-Bogoliubov (GHFB) with and without the angular momentum projection 
(AMP), since Tanaka {\it el al.} measured interaction cross sections 
$\sigma_{\rm I} (\approx \sigma_{\rm R})$  for $^{42-51}$Ca in RIKEN.  
As a review article on dynamical mean field approach, it is useful to see Ref.~\cite{Simenel:2012zc}.

\begin{figure}[H]
\centering
\vspace{0cm}
\includegraphics[width=0.5\textwidth]{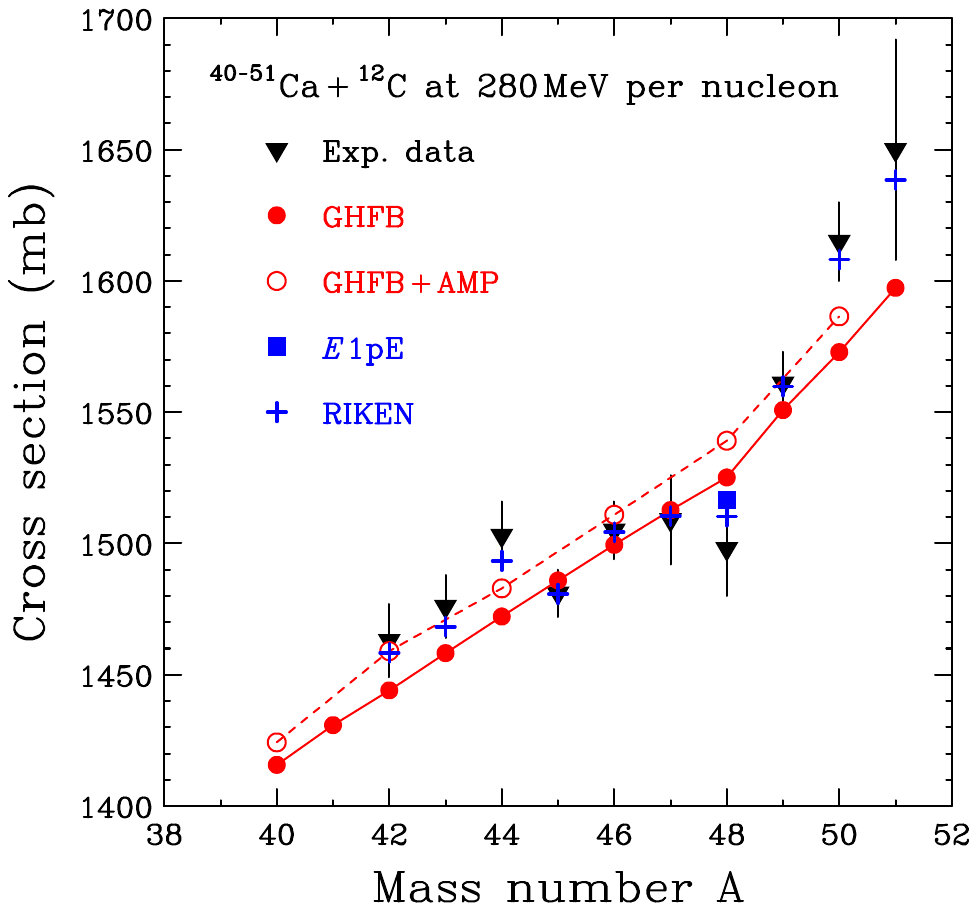}
\caption{
Mass-number dependence of $\sigma_{\rm R}$ for $^{42-51}$Ca+$^{12}$C scattering at  280~MeV per nucleon. 
The folding-model results with GHFB and GHFB+AMP densities are denoted by open and closed circles, 
respectively. 
The $\sigma_{\rm R}(E1{\rm pE})$ for $^{48}$Ca is shown by squares; see 
Sec.~\ref{$^{48}$Ca} for the definition of $\sigma_{\rm R}(E1{\rm pE})$. 
We scale the proton and neutron densities calculated with GHFB and GHFB+AMP so as to  
$r_{\rm p}({\rm scaling})=r_{\rm p}({\rm exp})$ and $r_{\rm n}({\rm scaling})=r_{\rm n}({\rm RIKEN})$. 
The results $\sigma_{\rm R}({\rm RIKEN})$ of the folding model  with the scaled densities are 
shown by symbol ``$+$''; see Tabel~\ref{Radii-exp} for $r_{\rm p}({\rm exp})$ and $r_{\rm n}({\rm RIKEN})$. 
The scaling equation is shown in Sec.~\ref{The scaling}.
Data on $\sigma_{\rm I}$ are taken from  Ref.~\cite{Tanaka:2019pdo} for $^{42-51}$Ca. 
}
\label{fig:$A$ dependence of-Ca+C}
\end{figure}

As shown in Fig. \ref{fig:$A$ dependence of-Ca+C}, the predicted  
$\sigma_{\rm R}$ results reproduce the data~\cite{Tanaka:2019pdo} in a $2\sigma$ level. 
This indicates that the  Kyushu $g$-matrix folding model with the GHFB and GHFB+AMP densities is good. 

Our purpose is to redetermine the $r_{\rm skin}$ from the $\sigma_{\rm I}$ with the Kyushu (chiral) 
$g$-matrix folding model. 
The Kyushu $g$-matrix folding model (the  GHFB and GHFB+AMP densities) is much better than the optical limit of the Glauber model (the Woos-Saxon densities). 

We scale the  GHFB and GHFB+AMP densities
so that the $\sigma_{\rm R}$ of the scaled densities can agree with the central values of $\sigma_{\rm I}$ 
under the condition that the proton radius of the scaled proton density 
equals the data~\cite{Angeli:2013epw} determined from the isotope shift  based on the electron scattering.

We explain our model in Sec.~\ref{Model} and our results in Sec.~\ref{Results}. 
Section \ref{Sec:discussions} is for discussions. 
Section \ref{Summary} is devoted to a summary.


\section{Model}
\label{Model}

\subsection{Folding model}

In the $g$-matrix folding model, the potential $U(\vR)$ consists 
of the direct and the exchange part defined in Ref. \cite{Sumi:2012fr}:
\bea
\label{eq:UD}
U^{\rm DR}(\vR) \hspace*{-0.15cm} &=& \hspace*{-0.15cm} 
\sum_{\mu,\nu}\int \rho^{\mu}_{\rm P}(\vrr_{\rm P}) 
            \rho^{\nu}_{\rm T}(\vrr_{\rm T})
           g^{\rm DR}_{\mu\nu}(s) d \vrr_{\rm P} d \vrr_{\rm T}, \\
\label{eq:UEX}
U^{\rm EX}(\vR) \hspace*{-0.15cm} &=& \hspace*{-0.15cm}\sum_{\mu,\nu} 
\int \rho^{\mu}_{\rm P}(\vrr_{\rm P},\vrr_{\rm P}-\vs)
\rho^{\nu}_{\rm T}(\vrr_{\rm T},\vrr_{\rm T}+\vs) \nonumber \\
            &&~~\hspace*{-0.5cm}\times g^{\rm EX}_{\mu\nu}(s) \exp{[-i\vK(\vR) \cdot \vs/M]}
            d \vrr_{\rm P} d \vrr_{\rm T},~~~~
            \label{U-EX}
\eea
where $\vs=\vrr_{\rm P}-\vrr_{\rm T}+\vR$ 
for the coordinate $\vR$ between a projectile (P) and a target (T). The coordinate 
$\vrr_{\rm P}$ 
($\vrr_{\rm T}$) denotes the location for the interacting nucleon 
measured from the center-of-mass of P (T). 
Each of $\mu$ and $\nu$ stands for the $z$-component
of isospin; 1/2 means neutron and $-$1/2 does proton.
The original form of $U^{\rm EX}$ is a non-local function of $\vR$,
but  it has been localized in Eq.~\eqref{U-EX}
with the local semi-classical approximation~\cite{Brieva-Rook} in which
P is assumed to propagate as a plane wave with
the local momentum $\hbar \vK(\vR)$ within a short range of the 
nucleon-nucleon interaction, where $M=A A_{\rm T}/(A +A_{\rm T})$
for the mass number $A$ ($A_{\rm T}$) of P (T).
The validity of this localization is shown in Ref.~\cite{Minomo:2009ds}.

The direct and exchange parts, $g^{\rm DR}_{\mu\nu}$ and 
$g^{\rm EX}_{\mu\nu}$, of the $g$ matrix are described by
\begin{align}
&\hspace*{0.5cm} g_{\mu\nu}^{\rm DR}(s) \nonumber \\ 
&=
\begin{cases}
\displaystyle{\frac{1}{4} \sum_S} \hat{S}^2 t_{\mu\nu}^{S1}
 (s) \hspace*{0.42cm} ; \hspace*{0.2cm} 
 {\rm for} \hspace*{0.1cm} \mu+\nu = \pm 1 
 \vspace*{0.2cm}\\
\displaystyle{\frac{1}{8} \sum_{S,T}} 
\hat{S}^2 t_{\mu\nu}^{ST}(s), 
\hspace*{0.2cm} ; \hspace*{0.2cm} 
{\rm for} \hspace*{0.1cm} \mu+\nu = 0 
\end{cases}
\\
&\hspace*{0.5cm}
g_{\mu\nu}^{\rm EX}(s) \nonumber \\
&=
\begin{cases}
\displaystyle{\frac{1}{4} \sum_S} (-1)^{S+1} 
\hat{S}^2 t_{\mu\nu}^{S1} (s) 
\hspace*{0.34cm} ; \hspace*{0.2cm} 
{\rm for} \hspace*{0.1cm} \mu+\nu = \pm 1 \vspace*{0.2cm}\\
\displaystyle{\frac{1}{8} \sum_{S,T}} (-1)^{S+T} 
\hat{S}^2 t_{\mu\nu}^{ST}(s) 
\hspace*{0.2cm} ; \hspace*{0.2cm}
{\rm for} \hspace*{0.1cm} \mu+\nu = 0 ~~~~~
\end{cases}
\end{align}
where  
the $g_{\mu\nu}^{ST}$  are 
the spin-isospin ($S$-$T$) components of the $g$-matrix interaction and $\hat{S} = {\sqrt {2S+1}}$.
As a way of the center-of-mass (cm) corrections in the proton and neutron densities,
we take the method  of Ref.~\cite{Sumi:2012fr}, since it is very simple. 
As for $^{12}$C, we use a phenomenological density of Ref.~\cite{C12-density}. 
As for Ca isotopes, we take the  densities scaled from the GHFB and GHFB+AMP densities.

\subsection{GHFB and GHFB+AMP}

In GHFB+AMP, the total wave function  $| \Psi^I_{M} \rangle$ with the AMP is defined by 
\begin{equation}
 | \Psi^I_{M} \rangle =
 \sum_{K, n=1}^{N+1} g^I_{K n} \hat P^I_{MK}|\Phi_n \rangle ,
\label{eq:prjc}
\end{equation}
where $\hat P^I_{MK}$ is the angular-momentum-projector and the 
$|\Phi_n \rangle$ for $n=1,2,\cdots,N+1$ are mean-field (GHFB) states, 
where $N$ is the number  of the  states.  
The coefficients $g^I_{K n}$ are determined
by solving the following Hill-Wheeler equation,
\begin{equation}
 \sum_{K^\prime n^\prime }{\cal H}^I_{Kn,K^\prime n^\prime }\ g^I_{K^\prime n^\prime } =
 E_I\,
 \sum_{K^\prime n^\prime }{\cal N}^I_{Kn,K^\prime n^\prime }\ g^I_{K^\prime n^\prime },
\end{equation}
with the Hamiltonian and norm kernels defined by
\begin{equation}
 \left\{ \begin{array}{c}
   {\cal H}^I_{Kn,K^\prime n^\prime } \\ {\cal N}^I_{Kn,K^\prime n^\prime } \end{array}
 \right\} = \langle \Phi_n |
 \left\{ \begin{array}{c}
   \hat{H} \\ 1 \end{array}
 \right\} \hat{P}_{KK^\prime }^I | \Phi_{n'} \rangle.
\end{equation}

For odd nuclei,  we have to put a quasi-particle in a level, but the number of the blocking states 
are quite large. It is difficult to solve  the Hill-Wheeler equation with large $N$. 
Furthermore, we have to confirm that the resulting  $| \Psi^I_{M} \rangle$ converges with respect to 
increasing $N$ for any set of two deformations  $\beta$ and $\gamma$. This procedure is quite time-consuming. For this reason, it is not feasible to perform the AMP for  odd nuclei. 
As for GHFB, we consider the one-quasiparticle state that yields the lowest energy, so that 
we do not have to solve the Hill-Wheeler equation. However, it is not easy  to find the values of 
$\beta$ and $\gamma$ at which the energy becomes minimum in the $\beta$-$\gamma$ plane.

For even nuclei, there is no blocking state in the Hill-Wheeler equation. 
We can thus consider GHFB+AMP. 
However,  we have to find the value of $\beta$ at 
which the ground-state energy becomes minimum.  
In this step, the AMP has to be performed for any $\beta$, so that the Hill-Wheeler calculation is still heavy. 
In fact, the AMP is not taken for most of mean field calculations; see for example Ref.~\cite{HP:AMEDEE}.  
The reason why we do not take into account $\gamma$ deformation is 
that the deformation does not affect  $\sigma_{\rm R}$~\cite{Sumi:2012fr}. 

\subsection{The scaling of the GHFB and GHFB+AMP densities}
\label{The scaling}

We explain the scaling of original density $\rho(\vrr)_{\rm original}$.  
We can obtain the scaled density $\rho_{\rm scaling}(\vrr)$ from the original one as
\bea
\rho_{\rm scaling}(\vrr)=\frac{1}{\a^3}\rho_{\rm original}(\vrr/\a)
\eea
with a scaling factor
\bea
\a=\sqrt{ \frac{\langle \vrr^2 \rangle_{\rm scaling}}{\langle \vrr^2 \rangle_{\rm original}}} .
\eea
For later convenience, we refer to the proton (neutron) radius of 
the scaled density as $r_{\rm p}({\rm scaling})$ ( $r_{\rm n}({\rm scaling})$).

\section{Results}
\label{Results}

\subsection{$^{42-51}$Ca}

Table \ref{Radii for Ca isotopes. } show theoretical radii determined with GHFB and GFHB+AMP for $^{39-64}$Ca. 
Effects of the AMP are small for radii. 

\begin{table}[htb]
\begin{center}
\caption
{Theoretical radii for Ca isotopes. 
The superscript ``AMP'' stands for the results of GHFB+AMP, and 
no 
superscript corresponds to those of GHFB.  }
 \begin{tabular}{ccccccccc}
  \hline
$A$ & $r_n^{\rm AMP}$ & $r_p^{\rm AMP}$ & $r_m^{\rm AMP}$ & $r_{\rm skin}^{\rm AMP}$ &
$r_n$ & $r_p$ & $r_m$ & $r_{\rm skin}$ \\
  \hline
  39 & & & & &3.320  & 3.381  & 3.351  & -0.061  \cr
  40 & 3.366 & 3.412 & 3.389 & -0.046 & 3.349 & 3.393 & 3.371  & -0.044 \cr
  41 &  &  &  &  & 3.387 & 3.397 & 3.392 & -0.010  \cr
  42 & 3.451 & 3.424 & 3.438 & 0.026 & 3.417 & 3.401 & 3.409  & -0.010 \cr
  43 &  &  &  &  & 3.448 & 3.405 & 3.428 & 0.043  \cr
  44 & 3.501 & 3.426 & 3.467  & 0.075 & 3.477 & 3.410 & 3.447 &  0.067 \cr
  45 &  &  &  &  & 3.504 & 3.414 & 3.465 & 0.090  \cr
  46 & 3.555 & 3.436 & 3.504 & 0.118 & 3.530 & 3.420 & 3.483 &  0.110 \cr
  47 &  &  &  &  & 3.554 & 3.424 & 3.499 & 0.130  \cr
  48 & 3.604 & 3.445 & 3.539 & 0.159 & 3.576 & 3.428 & 3.515 & 0.148 \cr
  49 &  &  &  &  & 3.621 & 3.440 & 3.548 & 0.181  \cr
  50 & 3.687 & 3.469 & 3.601 & 0.218 & 3.658 & 3.452 & 3.577 &  0.206 \cr
  51 &  &  &  &  & 3.698 & 3.462 & 3.607 & 0.236 \cr
  52 & 3.760 & 3.490 & 3.659 & 0.270 & 3.734 & 3.475 & 3.659 &  0.270 \cr
  53 &  &  &  &  & 3.779 & 3.486 & 3.671 & 0.293 \cr
  54 & 3.840 & 3.524 & 3.726 & 0.316 & 3.817 & 3.507 & 3.705 &  0.310 \cr
  55 &  &  &  &  & 3.856 & 3.524 & 3.739 & 0.332  \cr
  56 & 3.913 & 3.557 & 3.790 & 0.357 & 3.891 & 3.541 & 3.770 &  0.350 \cr
  57 &  &  &  &  & 3.928 & 3.557 & 3.802 & 0.370  \cr
  58 & 3.977 & 3.588 & 3.847 & 0.389 & 3.958 & 3.575 & 3.830 &  0.383 \cr
  59 &  &  &  &  & 3.995 & 3.593 & 3.863 & 0.402  \cr
  60 & 4.043 & 3.611 & 3.904 & 0.432 & 4.020 & 3.608 & 3.888 &  0.412 \cr
  62 & 4.106 & 3.637 & 3.961 & 0.469 & 4.067 & 3.628 & 3.931 &  0.439 \cr
  64 & 4.153 & 3.658 & 4.005 & 0.494 & 4.113 & 3.648 & 3.974 &  0.465 \cr
  
  \hline
 \end{tabular}   
 \label{Radii for Ca isotopes. }
\end{center} \end{table}

As  proton and  neutron densities, we use GHFB for odd nuclei and GHFB+AMP for even nuclei, and 
scale the GHFB and GHFB+AMP densities so that the scaled proton and neutron radii may agree with 
$r_{\rm p}({\rm exp})$~\cite{Angeli:2013epw} of electron scattering and $r_{\rm n}({\rm RIKEN})$, respectively; 
namely $r_{\rm p}({\rm scaling})=r_{\rm p}({\rm exp})$ and $r_{\rm n}({\rm scaling})=r_{\rm n}({\rm RIKEN})$. 

Figure~\ref{fig:$A$ dependence of-Ca+C} shows mass-number ($A$) dependence of  $\sigma_{\rm R}$ for 
$^{42-51}$Ca scattering on a $^{12}$C target at  280~MeV per nucleon. 
The folding model with GHFB and GHFB+AMP densities  (open and closed circles) reproduce 
the data~\cite{Tanaka:2019pdo} in a 2$\sigma$ level, indicating that the folding model is reliable. 
This allows us to scale the proton and neutron densities calculated with GHFB and GHFB+AMP so as to  
$r_{\rm p}({\rm scaling})=r_{\rm p}({\rm exp})$ and $r_{\rm n}({\rm scaling})=r_{\rm n}({\rm RIKEN})$. 
The folding-model results ($+$) with the scaled densities  mentioned above slightly deviate 
the central values of $\sigma_{\rm I}$. The small deviation comes from the method taken.

Now we redetermine  $r_{\rm n}$, $r_{\rm m}$ and $r_{\rm skin}$ from the data~\cite{Tanaka:2019pdo} 
on $\sigma_{\rm I}$, using $r_{\rm p}({\rm exp})$~\cite{Angeli:2013epw} of electron scattering. 
For this purpose, we scale the proton and neutron  densities of GHFB and GHFB+AMP 
so that the $\sigma_{\rm R}$ calculated with the scaled densities may agree with 
the central values of $\sigma_{\rm I}$ 
under the condition that $r_{\rm p}({\rm scaling})=r_{\rm p}({\rm exp})$. 
The resulting values $r_{\rm n}(\sigma_{\rm I})$ and the $r_{\rm p}({\rm exp})$ 
yield $r_{\rm m}(\sigma_{\rm I})$ and $r_{\rm skin}(\sigma_{\rm I})$. Our results are tabulated in 
Table \ref{Radii-exp-ours}.

\begin{table}[htb]
\begin{center}
\caption
{Our values  on  $r_{\rm p}({\rm exp})$, $r_{\rm m}(\sigma_{\rm I})$, $r_{\rm n}(\sigma_{\rm I})$, 
$r_{\rm skin}(\sigma_{\rm I})$ for $^{42-51}$Ca. 
The $r_{\rm p}({\rm exp})$ are deduced from the charge radii~\cite{Angeli:2013epw}. 
The errors are taken from the original data of Ref.~\cite{Tanaka:2019pdo}.
 }
\begin{tabular}{ccccc}
\hline\hline
A & $r_{\rm p}({\rm exp})$~fm & $r_{\rm m}(\sigma_{\rm I})$~fm &  $r_{\rm n}(\sigma_{\rm I})$~fm & $r_{\rm skin}(\sigma_{\rm I})$~fm \\
\hline
42 & $3.411 \pm 0.003$ & $3.446 \pm 0.030$ & $3.477\pm 0.06$ & $0.066\pm 0.06$ \\
43 & 3.397 $\pm$ 0.003& 3.468 $\pm$ 0.029 & 3.529 $\pm$ 0.05 & 0.132 $\pm$ 0.05\\
44 & 3.424 $\pm$ 0.003 & 3.511 $\pm$ 0.030 & 3.582 $\pm$ 0.05 & 0.158 $\pm$ 0.05\\
45 & 3.401 $\pm$ 0.003 & 3.452 $\pm$ 0.026 & 3.493 $\pm$ 0.05 & 0.092 $\pm$ 0.05\\
46 & 3.401 $\pm$ 0.003 & 3.489 $\pm$ 0.026 & 3.555 $\pm$ 0.05 & 0.154 $\pm$ 0.05\\
47 & 3.384 $\pm$ 0.003 & 3.488 $\pm$ 0.034 & 3.563 $\pm$ 0.06  & 0.179 $\pm$ 0.06\\
48 & 3.385 $\pm$ 0.003 & 3.447 $\pm$ 0.035 & 3.490 $\pm$ 0.06 & 0.105 $\pm$ 0.06\\
49 & 3.400 $\pm$ 0.003 & 3.568 $\pm$ 0.028 & 3.679 $\pm$ 0.05 & 0.279 $\pm$ 0.05\\
50 & 3.429 $\pm$ 0.003 & 3.658 $\pm$ 0.031 & 3.803 $\pm$ 0.05  & 0.374 $\pm$ 0.05\\
51 & 3.445 $\pm$ 0.003 & 3.713 $\pm$ 0.066 & 3.877 $\pm$ 0.10 & 0.432 $\pm$ 0.10\\
\hline
\end{tabular}
 \label{Radii-exp-ours}
 \end{center} \end{table}

\subsection{$^{48}$Ca}
\label{$^{48}$Ca}

We consider  $r_{\rm skin}^{48}$, since $r_{\rm skin}^{48}$ is related to the slope parameter 
$L$ in neutron matter~\cite{Tagami:2020shn}. 
As a measurement on skin $r_{\rm skin}^{48}$, the high-resolution $E1$ polarizability 
experiment ($E1$pE)  was made for $^{48}$Ca~\cite{Birkhan:2016qkr} in RCNP. 
The result is 
\bea
r_{\rm skin}^{48}(E1{\rm pE}) =0.17 \pm 0.03=0.14-0.20~{\rm fm}.  
\label{Eq:skin-Ca48-E1}
\eea
For $r_{\rm skin}^{48}$, the  measurement is most reliable in the present stage. 
The central value 0.17~fm of 
Eq. \eqref{Eq:skin-Ca48-E1} yields  matter radius $r_{\rm m}(E1{\rm pE})=3.485$~fm and neutron radius 
$r_{\rm n}(E1{\rm pE})=3.555$~fm from proton radius $r_{\rm p}({\rm exp})=3.385$~fm evaluated 
with the isotope shift method based on the electron scattering~\cite{Angeli:2013epw}.  
We then scale the proton and neutron densities calculated with GHFB+AMP so as to 
reproduce $r_{\rm p}({\rm exp})$ and $r_{\rm n}(E1{\rm pE})$. 
In Fig.~\ref{fig:$A$ dependence of-Ca+C}, the $\sigma_{\rm R}(E1{\rm pE})$ calculated with 
the scaled densities is near  the upper bound of $\sigma_{\rm I}$.

\begin{figure}[H]
\centering
\vspace{0cm}
\includegraphics[width=0.5\textwidth]{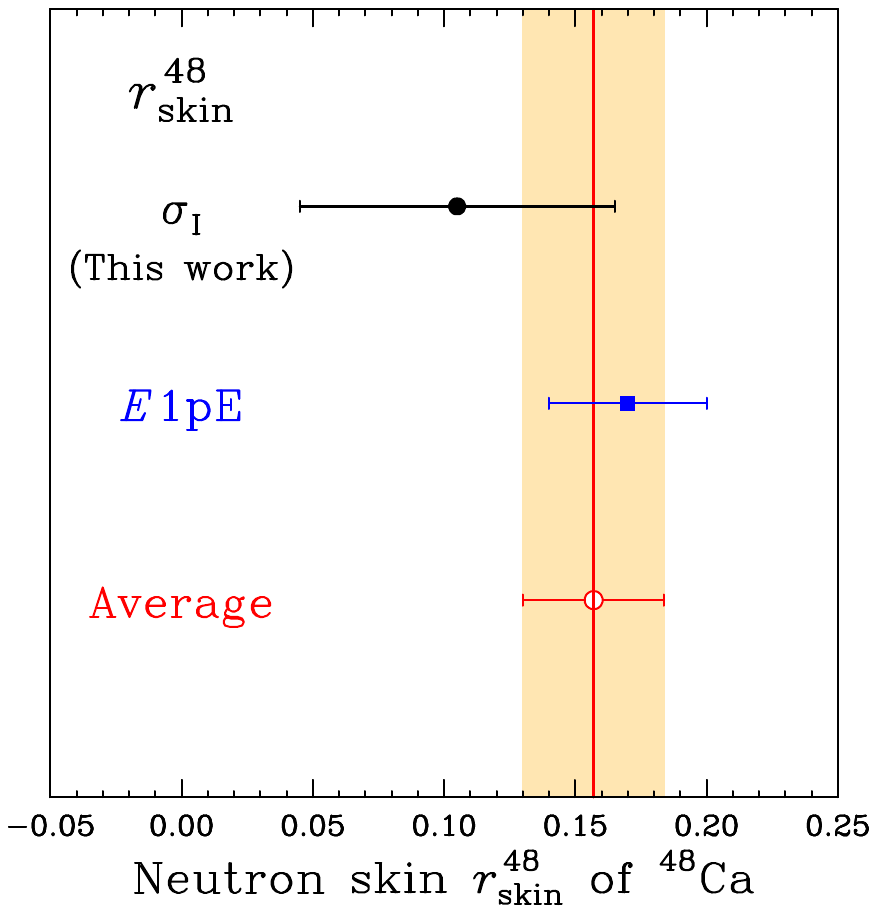}
\caption{
$r_{\rm skin}^{48}(\sigma_{\rm I})$, $r_{\rm skin}^{48}(E1{\rm pE})$, 
the weighted mean and its error for the two values. 
}
\label{final result}
\end{figure}

We take the weighted mean and its error for 
 $r_{\rm skin}^{48}(E1{\rm pE}) =0.17 \pm 0.03$~fm and our result 
 $r_{\rm skin}(\sigma_{\rm I})= 0.105 \pm 0.06$~fm. The final result is  
\bea
 r_{\rm skin}=0.157 \pm 0.027~{\rm fm}. 
 \label{eq:final result}
\eea
 Our final result is shown in Fig.~\ref{final result}, together with 
 $r_{\rm skin}^{48}(E1{\rm pE}) =0.17 \pm 0.03$~fm and our result 
 $r_{\rm skin}(\sigma_{\rm I})= 0.105 \pm 0.06$~fm.  
    
As an {\it ab initio} method for Ca isotopes,
we should consider the coupled-cluster method~\cite{Hagen:2013nca,Hagen:2015yea} with chiral interaction. 
Chiral interactions were constructed by 
two groups~\cite{Weinberg:1991um, Epelbaum:2008ga, Machleidt:2011zz}.  
The coupled-cluster result~\cite{Hagen:2015yea} 
\bea
r_{\rm skin}^{48}({\rm CC})=0.12 - 0.15~ {\rm fm}
\label{Eq:skin-Ca48-CC}
\eea 
is consistent with our final result of Eq.~\eqref{eq:final result}.

\section{Discussions}
\label{Sec:discussions}
Mass-number dependence $A$ of $\sigma_{I}$  has a kink at $A=48$. 
The data on $\alpha \equiv r_{m}E_{\rm B}/(A\hbar c)$ hardly depend on $A$, 
as shown in Table~\ref{table-a}; note that $E_{\rm B}$ is the binding energy of a nucleus. 
Here, the central values of data on $r_{m}$ and $E_{\rm B}$ are taken from  
Refs.~\cite{Tanaka:2019pdo,HP:NuDat 2.8}.
In fact, the deviation  of  $\alpha$ is  much smaller  than the average value; namely, 
\bea
\alpha=0.1535 (9) 
\label{Eq-a}
\eea
for  $^{42-51}$Ca. This indicates that  $r_m$ is in inverse proportion to $E_{\rm B}/A$ as an experimental result.

\begin{table}[H]
\begin{center}
\caption
{Numerical values of  $\alpha \equiv r_{m}E_{\rm B}/(A\hbar c)$, $r_{\rm m}(\sigma_{\rm I})$, 
$E_{\rm B}/A$ for $^{42-51}$Ca. 
The data $r_{\rm m}(\sigma_{\rm I})$ on $r_{\rm m}$ are taken from  Ref.~\cite{Tanaka:2019pdo}, 
and the data on $E_{\rm B}/A$ are  from Ref.~\cite{HP:NuDat 2.8}.
 }
\begin{tabular}{cccc}
\hline\hline
A & $r_{\rm m}({\rm RIKEN})$~fm &  $E_{\rm B}/A$~MeV &  $\alpha$ \\
\hline
42 & 3.437	&8.616563	&0.1501 \\
43	&3.453	&8.600663	&0.1505 \\
44	&3.492	&8.658175	&0.1532 \\
45	&3.452	&8.630545	&0.1510 \\
46	&3.487	&8.66898	      &0.1532 \\
47	&3.491	&8.63935	      &0.1528 \\
48	&3.471	&8.666686	&0.1524 \\
49	&3.565	&8.594844	&0.1553 \\
50	&3.645	&8.55016	      &0.1579 \\
51	&3.692	&8.476913	&0.1586 \\ 
\hline
\end{tabular}
 \label{table-a}
 \end{center} \end{table}


For $r_{\rm skin}^{48}$, the difference between $\sigma_{\rm R}(E1{\rm pE})$ and the central value of 
$\sigma_{\rm I}$ may come from that between reaction cross section and 
interaction cross section, i.e., $\sigma_{\rm R}(E1{\rm pE})-\sigma_{\rm I}=18.5$~mb. 
We assume that the difference $\sigma_{\rm R}(E1{\rm pE})-\sigma_{\rm I}=18.5$~mb for $^{48}$Ca  
is the same as for $^{42-51}$Ca.  The estimated $\sigma_{\rm R}$ is $\sigma_{\rm I}$+18.5~mb in which 
 the error on the estimated $\sigma_{\rm R}$ is $2.5$\% larger than the error $1.1$\% on $\sigma_{\rm I}$; 
as a good experiment on $\sigma_{\rm R}$, we can consider $^{9}$Be, $^{12}$C, $^{27}$Al+$^{12}$C scattering 
of Ref.~\cite{Takechi:2009zz} and the error is $2 \sim 3$\%. 
The figure \ref{fig:$A$ dependence of-Ca+C-2} is shown below. 

\begin{figure}[H]
\centering
\vspace{0cm}
\includegraphics[width=0.5\textwidth]{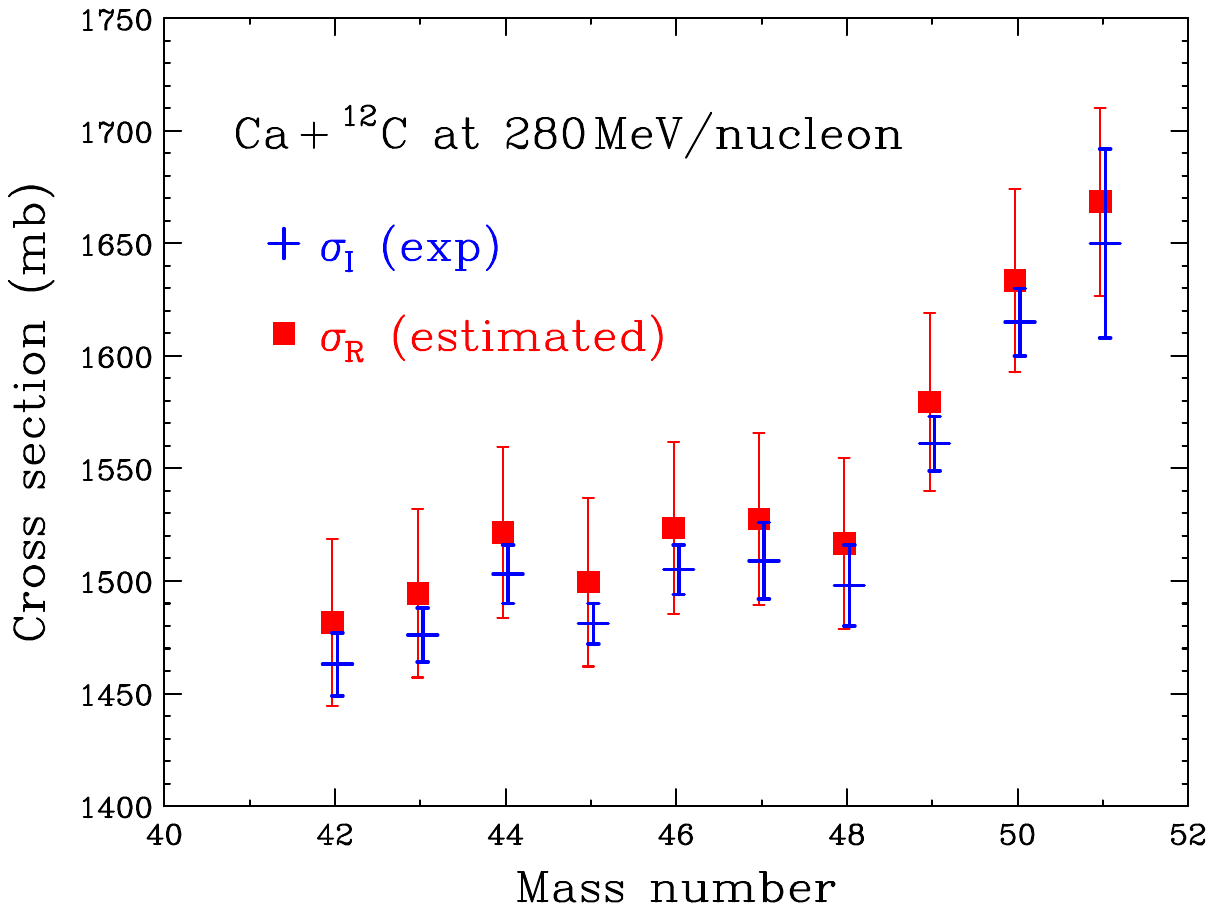}
\caption{
Mass-number dependence of $\sigma_{\rm R}$ and $\sigma_{\rm I}$ 
for $^{42-51}$Ca+$^{12}$C scattering at  280~MeV per nucleon. 
The estimated $\sigma_{\rm R}$ are shown by squares, while 
the data on $\sigma_{\rm I}$ are by the symbol ``+''. 
Data on $\sigma_{\rm I}$ are taken from  Ref.~\cite{Tanaka:2019pdo} for $^{42-51}$Ca. 
}
\label{fig:$A$ dependence of-Ca+C-2}
\end{figure}

Now we assume that $\sigma_{\rm R}=\sigma_{\rm I}({\rm RIKEN})$+18.5~mb 
as the data for $^{42-51}$Ca. 
Using the estimated $\sigma_{\rm R}$ instead of  $\sigma_{\rm I}$, we take the same procedure 
in order to obtain $r_{\rm skin}(\sigma_{\rm R})$. 
As shown in Table~\ref{skins-exp-estimated}, the resulting $r_{\rm skin}(\sigma_{\rm R})$ is 
almost the same as the $r_{\rm skin}({\rm RIKEN})$ of Ref.~\cite{Tanaka:2019pdo}, except for $^{48}$Ca.
As for $^{48}$Ca, our estimates value  0.170 $\pm$ 0.06~fm agrees with 
$r_{\rm skin}^{48}(E1{\rm pE}) =0.17 \pm 0.03~{\rm fm}$ of Ref.~\cite{Birkhan:2016qkr}.

\begin{table}[htb]
\begin{center}
\caption
{Numerical values 
of  estimated $r_{\rm skin}(\sigma_{\rm R})$, $r_{\rm skin}({\rm RIKEN})$~\cite{Tanaka:2019pdo} 
for $^{42-51}$Ca. 
The skins are shown in units of fm. 
 }
\begin{tabular}{ccc}
\hline\hline
A & $r_{\rm skin}(\sigma_{\rm R})$ & $r_{\rm skin}({\rm RIKEN})$ \\
\hline
42 &  $0.049\pm 0.06$ & $0.049\pm 0.06$ \\
43 &  0.104 $\pm$ 0.05 & 0.103 $\pm$ 0.05\\
44 &  0.124 $\pm$ 0.05 & 0.125 $\pm$ 0.05\\
45 &  0.091 $\pm$ 0.05 & 0.092 $\pm$ 0.05\\
46 &  0.151 $\pm$ 0.05 & 0.151 $\pm$ 0.05\\
47 &  0.184 $\pm$ 0.06  & 0.184 $\pm$ 0.06\\
48 &  0.170 $\pm$ 0.06 & 0.146 $\pm$ 0.06\\
49 &  0.275 $\pm$ 0.05 & 0.275 $\pm$ 0.05\\
50 &  0.353 $\pm$ 0.05  & 0.353 $\pm$ 0.05\\
51 &  0.398 $\pm$ 0.10 & 0.399 $\pm$ 0.10\\
\hline
\end{tabular}
 \label{skins-exp-estimated}
 \end{center} \end{table}

\section{Summary}
\label{Summary}

Recently, Tanaka {\it el al.} measured $\sigma_{\rm I}$ in RIKEN for $^{42-51}$Ca+ $^{12}$C 
scattering at 280~MeV per nucleon, and 
determined neutron skins $r_{\rm skin}$ for $^{42-51}$Ca from the $\sigma_{\rm I}$, 
using the optical limit of the Glauber model with the Woos-Saxon densities~\cite{Tanaka:2019pdo}. 
We redetermine   $r_{\rm skin}$,  $r_{\rm m}$, $r_{\rm n}$ for $^{42-51}$Ca, 
using the Kyushu folding model with the proton and neutron densities scaled from 
the GHFB and GHFB+AMP densities.

The $\sigma_{\rm R}$ calculated with the GHFB and GHFB+AMP densities 
almost reproduce the data~\cite{Tanaka:2019pdo} on $\sigma_{\rm I}$. 
This allows us to determine $r_{\rm skin}$ from the central values of $\sigma_{\rm I}$
by scaling the proton and neutron densities. The $r_{\rm skin}$ thus determined are close 
to the original ones of Ref.~\cite{Tanaka:2019pdo}, except for $r_{\rm skin}^{48}$; see 
Table \ref{Radii-exp} for the original values and Table \ref{Radii-exp-ours} for ours.  
The $r_{\rm skin}$ thus determined  are close to the original results 
$r_{\rm skin}^{42-51}({\rm RIKEN})$, 
except for $^{48}$Ca.  
Our experimental values on $r_{\rm m}$, $r_{\rm n}$, $r_{\rm skin}$ for $^{42-51}$Ca are summarized 
in Table \ref{Radii-exp-ours}. 

For $^{48}$Ca, our value is $r_{\rm skin}^{48}(\sigma_{\rm I})= 0.105 \pm 0.06$~fm, while  
Birkhan {\it et. al}. determined $r_{\rm skin}^{48}(E1{\rm pE}) =0.17 \pm 0.03$~fm~\cite{Birkhan:2016qkr} 
from the high-resolution $E1$ polarizability experiment (E1{\rm pE}).
We then take the weighted mean and its error 
for  the two values. The resulting value  $r_{\rm skin}^{48}=0.157 \pm 0.027$~fm is our final value 
for $^{48}$Ca. The value is related to CREX that is ongoing.

\section*{Acknowledgements}
We thank Dr. Tanaka and Prof. Fukuda for providing the data  
and helpful comments.  M. Y. thanks Dr. M. Toyokawa heartily. 




\begin{thebibliography}{19}
\expandafter\ifx\csname natexlab\endcsname\relax\def\natexlab#1{#1}\fi
\expandafter\ifx\csname bibnamefont\endcsname\relax
  \def\bibnamefont#1{#1}\fi
\expandafter\ifx\csname bibfnamefont\endcsname\relax
  \def\bibfnamefont#1{#1}\fi
\expandafter\ifx\csname citenamefont\endcsname\relax
  \def\citenamefont#1{#1}\fi
\expandafter\ifx\csname url\endcsname\relax
  \def\url#1{\texttt{#1}}\fi
\expandafter\ifx\csname urlprefix\endcsname\relax\def\urlprefix{URL }\fi
\providecommand{\bibinfo}[2]{#2}
\providecommand{\eprint}[2][]{\url{#2}}


\bibitem{Tanaka:2019pdo} 
  M.~Tanaka {\it et al.},
  Phys.\ Rev.\ Lett.\  {\bf 124}, 102501 (2020).
  [arXiv:1911.05262 [nucl-ex]].

\bibitem{Brieva-Rook}
F.~A.~Brieva and J.~R.~Rook, Nucl. Phys. A~{\bf 291}, 299 (1977);
{\it ibid.}~291, 317 (1977); {\it ibid.}~297, 206 (1978).

\bibitem{Amos}
K.~Amos, P.~J.~Dortmans, H.~V.~von Geramb, 
S.~Karataglidis, and J.~Raynal, 
in \textit{Advances in Nuclear Physics}, edited by
J.~W.~Negele and E.~Vogt(Plenum, New York, 2000) Vol.~25, p. 275.

\bibitem{CEG07}
T.~Furumoto, Y.~Sakuragi, and Y.~Yamamoto, 
Phys. Rev. C~{\bf 78}, 044610 (2008). 

\bibitem{Minomo:2011bb} 
  K.~Minomo, T.~Sumi, M.~Kimura, K.~Ogata, Y.~R.~Shimizu and M.~Yahiro,
  Phys.\ Rev.\ Lett.\  {\bf 108}, 052503 (2012),
  [arXiv:1110.3867 [nucl-th]].
  
\bibitem{Toyokawa:2013uua} 
  M.~Toyokawa, K.~Minomo and M.~Yahiro,
  Phys.\ Rev.\ C {\bf 88}, no. 5, 054602 (2013), 
  [arXiv:1304.7884 [nucl-th]].

\bibitem{Toyokawa:2014yma} 
  M.~Toyokawa, K.~Minomo, M.~Kohno and M.~Yahiro,
  J.\ Phys.\ G {\bf 42}, no. 2, 025104 (2015), 
  Erratum: [J.\ Phys.\ G {\bf 44}, no. 7, 079502 (2017)]
  [arXiv:1404.6895 [nucl-th]].

\bibitem{Toyokawa:2015zxa} 
  M.~Toyokawa, M.~Yahiro, T.~Matsumoto, K.~Minomo, K.~Ogata and M.~Kohno,
  Phys.\ Rev.\ C {\bf 92}, no. 2, 024618 (2015),
  Erratum: [Phys.\ Rev.\ C {\bf 96}, no. 5, 059905 (2017)],
  [arXiv:1507.02807 [nucl-th]].

\bibitem{Toyokawa:2017pdd} 
  M.~Toyokawa, M.~Yahiro, T.~Matsumoto and M.~Kohno,
  PTEP {\bf 2018}, 023D03 (2018), 
  [arXiv:1712.07033 [nucl-th]]. 
  See http://www.nt.phys.kyushu-u.ac.jp/english/gmatrix.html for Kyushu $g$-matrix. 
  
\bibitem{Sumi:2012fr} 
  T.~Sumi, K.~Minomo, S.~Tagami, M.~Kimura, T.~Matsumoto, K.~Ogata, Y.~R.~Shimizu and M.~Yahiro,
  Phys.\ Rev.\ C {\bf 85}, 064613 (2012),
  [arXiv:1201.2497 [nucl-th]].
  
  

\bibitem{Egashira:2014zda} 
  K.~Egashira, K.~Minomo, M.~Toyokawa, T.~Matsumoto and M.~Yahiro,
  Phys.\ Rev.\ C {\bf 89}, 064611 (2014).
  [arXiv:1404.2735 [nucl-th]].

 

\bibitem{Watanabe:2014zea} 
  S.~Watanabe {\it et al.},
  Phys.\ Rev.\ C {\bf 89}, no. 4, 044610 (2014),
  [arXiv:1404.2373 [nucl-th]].
  
  

  
\bibitem{Koh13}
M.~Kohno, 
Phys. Rev. C~{\bf 88}, 064005 (2013). \\
M.~Kohno, Phys. Rev. C {\bf 96}, 059903(E) (2017). 



\bibitem{von-Geramb-1991} 
H.~V.~von Geramb, K.~Amos, L.~Berge, S.~Br\"autigam, H.~Kohlhoff and 
A.~Ingemarsson, 
Phys. Rev. C~{\bf 44}, 73 (1991). 

\bibitem{Amos-1994}
P.~J.~Dortmans and K.~Amos, 
Phys. Rev. C~{\bf 49}, 1309 (1994). 

  
\bibitem{Tagami:2019svt} 
  S.~Tagami, M.~Tanaka, M.~Takechi, M.~Fukuda and M.~Yahiro,
  Phys.\ Rev.\ C {\bf 101}, no. 1, 014620 (2020), 
  [arXiv:1911.05417 [nucl-th]].

\bibitem{Simenel:2012zc}
C.~Simenel,
Lect. Notes Phys. \textbf{875}, 95-145 (2014)
doi:10.1007/978-3-319-01077-9\_4
[arXiv:1211.2387 [nucl-th]].

\bibitem{Angeli:2013epw} 
  I.~Angeli and K.~P.~Marinova,
  Atom.\ Data Nucl.\ Data Tabel.\  {\bf 99}, 69 (2013). 
  
\bibitem{Minomo:2009ds}
  K.~Minomo, K.~Ogata, M.~Kohno, Y.~R.~Shimizu, and M.~Yahiro,
  J.\ Phys.\ G {\bf 37}, 085011 (2010)
  [arXiv:0911.1184 [nucl-th]].



\bibitem{C12-density}
H. de Vries, C. W. de Jager, and C. de Vries,
At. Data Nucl. Data Tables \textbf{36}, 495 (1987).  
  

  


\bibitem{HP:AMEDEE}
S. Hilaire and M. Girod, Hartree-Fock-Bogoliubov results based on the Gogny force; http://www-phynu.cea.fr/science-en-ligne/carte-potentiels-microscopiques/carte-potentiel-nucleaire-eng.htm.    


  
  


\bibitem{Tagami:2020shn}
S.~Tagami, N.~Yasutake, M.~Fukuda and M.~Yahiro,
[arXiv:2003.06168 [nucl-th]].



\bibitem{Birkhan:2016qkr} 
  J.~Birkhan {\it et al.},
  Phys.\ Rev.\ Lett.\  {\bf 118}, no. 25, 252501 (2017),
  [arXiv:1611.07072 [nucl-ex]].
  








  
\bibitem{Hagen:2015yea} 
  G.~Hagen {\it et al.},
  Nature Phys.\  {\bf 12}, 186 (2015), 
  [arXiv:1509.07169 [nucl-th]].
  
\bibitem{Hagen:2013nca} 
  G.~Hagen, T.~Papenbrock, M.~Hjorth-Jensen and D.~J.~Dean,
  Rept.\ Prog.\ Phys.\  {\bf 77}, 096302 (2014), 
  [arXiv:1312.7872 [nucl-th]].


\bibitem{Weinberg:1991um} 
  S.~Weinberg,
  Nucl.\ Phys.\ B {\bf 363}, 3 (1991).

\bibitem{Epelbaum:2008ga} 
  E.~Epelbaum, H.~W.~Hammer and U.~G.~Meissner,
  Rev.\ Mod.\ Phys.\  {\bf 81}, 1773 (2009).
  [arXiv:0811.1338 [nucl-th]].


\bibitem{Machleidt:2011zz} 
  R.~Machleidt and D.~R.~Entem,
  Phys.\ Rept.\  {\bf 503}, 1 (2011), 
  [arXiv:1105.2919 [nucl-th]].
  
\bibitem{HP:NuDat 2.8} 
{the National Nuclear Data Center, NuDat 2.8; 
  https://www.nndc.bnl.gov/nudat2/}. 
  


\bibitem{Takechi:2009zz}
M.~Takechi, M.~Fukuda, M.~Mihara, K.~Tanaka, T.~Chinda, T.~Matsumasa, M.~Nishimoto, R.~Matsumiya, Y.~Nakashima and H.~Matsubara, \textit{et al.}
Phys. Rev. C \textbf{79} (2009), 061601
doi:10.1103/PhysRevC.79.061601









\end{thebibliography}
\end{document}